\documentclass[12pt,a4paper]{article}

\usepackage[numbers]{natbib}
\usepackage{amsfonts,amsmath,amssymb,amsthm}
\usepackage{cleveref}
\usepackage{url}
\usepackage{graphicx}
\usepackage{blindtext,titlefoot}

\newcommand{\etal}{\emph{et al.}\ }

\theoremstyle{definition}
\newtheorem{definition}{Definition}[section]

\title{Reliability analysis of general phased mission systems with a new survival signature}

\author{Xianzhen~Huang \\ {\normalsize School of  Mechanical Engineering and Automation,} \\\vspace{10pt} {\normalsize Northeastern University, China} \\ Louis~J.~M.~Aslett \\\vspace{10pt} {\normalsize Department of Mathematical Sciences, Durham University, UK} \\ Frank~P.~A.~Coolen \\ {\normalsize Department of Mathematical Sciences, Durham University, UK}}

\date{}

\begin{document}

\maketitle\unmarkedfntext{\textit{Email addresses:} \texttt{xzhhuang83@gmail.com} (X.~Huang), \texttt{louis.aslett@durham.ac.uk} (L.J.M.~Aslett), \texttt{frank.coolen@durham.ac.uk} (F.P.A.~Coolen)}

\vspace{-24pt}
\begin{abstract}
It is often difficult for a phased mission system (PMS) to be highly reliable, because this entails achieving high reliability in every phase of operation.  Consequently, reliability analysis of such systems is of critical importance.  However, efficient and interpretable analysis of PMSs enabling general component lifetime distributions, arbitrary structures, and the possibility that components skip phases has been an open problem.

In this paper, we show that the survival signature can be used for reliability analysis of PMSs with similar types of component in each phase, providing an alternative to the existing limited approaches in the literature.  We then develop new methodology addressing the full range of challenges above.  The new method retains the attractive survival signature property of separating the system structure from the component lifetime distributions, simplifying computation, insight into, and inference for system reliability.

\end{abstract}


\section{Introduction}

A phased mission system (PMS) is one that performs several different tasks or functions in sequence. The periods in which each of these successive tasks or functions takes place are known as phases \citep{xing2008reliability,la2004phased}.  Examples of PMSs can be found in many practical applications, such as electric power systems, aerospace systems, weapon systems and computer systems. A typical example of a PMS is the monitoring system in a satellite-launching mission with three phases: launch, separation, and orbiting.

A PMS is considered to be functioning if all of its phases are completed without failure, and failed if failure occurs in any phase. Therefore, the reliability of a PMS with $N$ phases is the probability that it operates successfully in all of its phases:
\begin{equation}
  R_S = \mathbb{P}(\mbox{Phase 1 works} \cap \mbox{Phase 2 works} \cap \dots \cap \mbox{Phase $N$ works})
  \label{eq:allphases}  
\end{equation}

The calculation of the reliability of a PMS is more complex than that of a single phase system, because the structure of the system varies between phases and the component failures in different phases are mutually dependent \citep{xing2008reliability}.

Over the past few decades, there have been extensive research efforts to analyze PMS reliability.  Generally, there are two classes of models to address such scenarios: state space oriented models \citep{kim1994phased,chew2008phased,lu2014reliability,wang2017competing} and combinatorial methods \citep{xing2015binary,ma1999algorithm,la2004phased,zang1999bdd,tang2006bdd,mo2009variable,reed2011improved,xing2007reliability,xing2013bdd}. The main idea of state space oriented models is to construct Markov chains and/or Petri nets to represent the system behaviour, since these provide flexible and powerful options for modelling complex dependencies among system components. However, the cardinality of the state space can become exponentially large as the number of components increases.  The remaining approaches exploit combinatorial methods, Boolean algebra and various forms of decision diagrams for reliability analysis of PMSs.

In particular, in recent years the Binary Decision Diagram (BDD) --- a combinatorial method --- has become more widely used in reliability analysis of PMSs due to its computationally efficient and compact representation of the structure function compared with other methods. Zang \etal \citep{zang1999bdd} first used the BDD method to analyze the reliability of PMSs. Tang \etal \citep{tang2006bdd} developed a new BDD-based algorithm for reliability analysis of PMSs with multimode failures. Mo \citep{mo2009variable} and Reed \etal \citep{reed2011improved} improved the efficiency of Tang's method by proposing a heuristic selection strategy and reducing the BDD size, respectively. Xing \etal \citep{xing2007reliability,xing2013bdd} and Levitin \etal \citep{levitin2013reliability} proposed BDD based methods for the reliability evaluation of PMSs with common-cause failures and propagated failures.  Wang \etal \citep{wang2007reliability} and Lu \etal \citep{lu2015reliability} studied modular methods for reliability analysis of PMSs with repairable components, by combining BDDs with state-enumeration methods.

While the BDD method has been shown to be a very efficient combinatorial method, it is still difficult to analyze large systems without considerable computational expense \citep{xing2008reliability,reed2011improved}.  In this paper, we propose a combinatorial analytical approach providing a new survival signature methodology for reliability analysis of PMSs.  This paper is organized as follows: \cref{sec:PMS} gives a brief background on PMSs; \cref{sec:survsig} first shows how the standard survival signature can be used to evaluate PMSs with similar component types in each phase, before providing a novel methodology which facilitates heterogeneity of components across the phases.  \Cref{sec:examples} presents illustrative examples showing numerical agreement with existing literature, but where the full benefits of the interpretability of survival signatures is now available due to this work.  Finally, \cref{sec:conclusion} presents some conclusions ideas for future work.

\section{Phased mission systems}
\label{sec:PMS}

\begin{figure}
  \centering
  \includegraphics[width=0.8\textwidth]{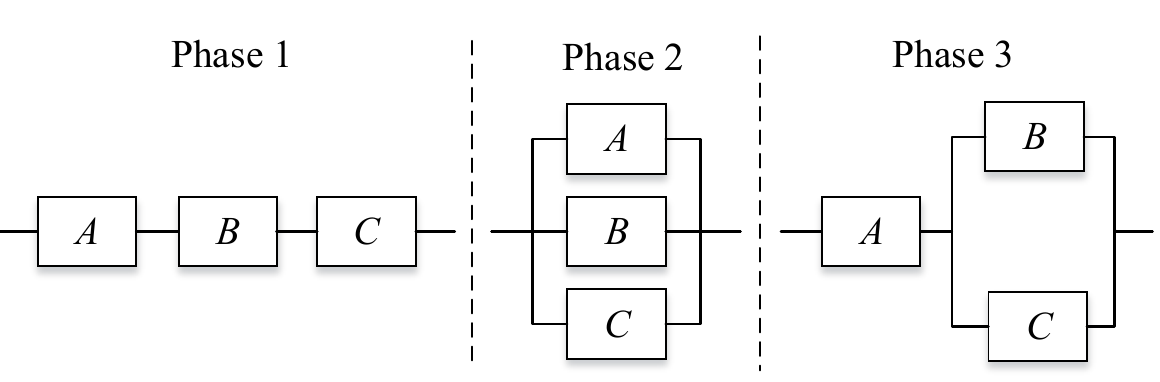}
  \caption{A PMS with similar components in each phase.}
  \label{fig:pms1}
\end{figure}

\Cref{fig:pms1} shows a simple system that performs a series of functions or tasks which are carried out over consecutive periods of time to achieve a certain overall goal (or `mission').  Such a system --- where the structure (and possibly operating environment) of the system changes over time --- is known as a Phased Mission System (PMS), with each period of operation being referred to as a `phase'. Each phase therefore corresponds to one structural configuration and components in different phases are taken to be mutually dependent.

Let us consider a system consisting of $N \ge 2$ phases, with $n_i$ components in phase $i \in \{1, \dots, N\}$.  The binary state indicator variable $X_{ij}, j \in \{1, \dots, n_i\}$ denotes the operational status of the $j$th component in phase $i$:
\[ X_{ij} = \begin{cases}
1 & \mbox{if component $j$ works for all of phase $i$} \\
0 & \mbox{if component $j$ fails before the end of phase $i$}
\end{cases} \]

The vectors $\mathbf{X}_i = (X_{i1}, \dots, X_{in_i}), i \in \{1, \dots, N\}$, represent the states of all components in the $i$th phase and the full vector $\mathbf{X} = (\mathbf{X}_1, \dots, \mathbf{X}_N) = (X_{11}, \dots, X_{1n_1}, \dots, X_{N1}, \dots, X_{Nn_N})$ represents the states of all components during the full mission.

The state of the system in each phase is also a binary random variable, which is completely determined by the states of the components in that phase.  Let $\phi_i$ represent the system state in the $i$th phase, that is:
\[ \phi_i = \varphi_i(\mathbf{X}_i) = \varphi_i(X_{i1}, \dots, X_{in_i}) \]
where $\varphi_i(\cdot)$ is the structure function of the system design in phase $i$.  The structure function evaluates to $\phi_i = 1$ if the system functions for state vector $\mathbf{X}_i$, and $\phi_i = 0$ if not.

Similarly, the structure function of the full PMS (that is, the operational state of the system across \emph{all} phases) is also a binary random variable, which is completely determined by the states of all the components in the PMS
\begin{equation}
  \phi_S = \varphi_S(\mathbf{X}) \triangleq \prod_{i=1}^N \varphi_i(X_{i1}, \dots, X_{in_i}) 
  \label{eq:strfnpms}
\end{equation}

The structure function as shown in \cref{eq:strfnpms} is again a Boolean function which is derived from the truth table of the structure functions for each phase of operation.  The truth tables depend uniquely on the system configurations and simply provide a means of tabulating all the possible combinational states of each component to realise the operational state of the system in each case. The state vectors for which $\varphi_S(\mathbf{X})=1$ provide a logical expression for the functioning of the system, while the states when $\varphi_S(\mathbf{X})=0$ provide a logical expression for the failure of the system. It should be noted that, unlike non-PMSs, there exist impossible combinations of states which should be deleted from the truth table when performing a reliability analysis.  For example, if both the system and its components are non-repairable during the mission, then if a component is failed in a certain phase it cannot be working in subsequent phases.

Finally, if all phases are completed successfully, the mission is a success, that is:
\[ \phi_S = \prod_{i=1}^N \phi_i = 1 \iff \phi_i = 1 \ \forall\,i \]

\section{Survival signature}
\label{sec:survsig}

For larger systems, working with the full structure function can be complicated and as the system size grows it becomes hard to intuit anything meaningful from the particular algebraic form it takes.  In particular, one may be able to summarize the structure function when it consists of exchangeable components of one or more types \citep{samaniego2007system,coolen2013generalizing,coolen2014nonparametric}.

Recently, the concept of the survival signature has attracted substantial attention, because it provides such a summary which enables insight into the system design even for large numbers of components of differing types.  Coolen and Coolen-Maturi \citep{coolen2013generalizing} first introduced the survival signature, using it to analyze complex systems consisting of multiple types of component. Subsequently, \citep{coolen2014nonparametric,coolen2015predictive,Aslett2015} presented the use of the survival signature in an inferential setting, with nonparametric predictive inference and Bayesian posterior predictive inference respectively, and \citep{feng2016imprecise} presented methods for analyzing imprecise system reliability using the survival signature. Patelli \etal \citep{patelli2017simulation} developed a survival signature-based simulation method  to calculate the reliability of large and complex systems and \citep{Aslett2017} presents a simulation method which can be used if the dependency structure is too complex for a survival signature approach.  Walter \etal \citep{walter2017condition} proposed a new condition-based maintenance policy for complex systems using the survival signature.  Moreover, Eryilmaz \etal \citep{eryilmaz2016generalizing} generalized the survival signature to multi-state systems.

Efficient computation of the survival signature was addressed by Reed \citep{reed2017efficient}, using reduced order binary decision diagrams (ROBDDs).  The survival signature of a system can be easily computed by specifying the reliability block diagram as a simple graph by using the \texttt{ReliabilityTheory} R package \citep{Aslett2012}.

In this section, the survival signature is first shown to apply directly to full mission-length PMSs where there is a single component type in each phase.  Thereafter, an extension is presented which enables heterogeneity of component types across phases, providing novel methodology for reliability analysis of PMSs.

\subsection{PMSs with similar components in each phase}
\label{sec:pms.same}

We consider a system with $N \ge 2$ phases, with $n$ components in each phase (e.g.\ the PMS as shown in \cref{fig:pms1}), and let phase $i \in \{1, \dots, N\}$ run from time $\tau_i$ to time $\tau_{i+1}$ with $\tau_1 \triangleq 0$ and $\tau_i < \tau_{i+1} \ \forall\,i$.  Thus the full mission time is denoted $\tau_{N+1}$.

We assume that the random failure times of components in the same phase are fully independent, and in addition that the components are exchangeable. Let $\Phi(l_1, \dots, l_N)$ denote the probability that the PMS functions by the end of the mission given that precisely $l_i, i \in \{1, \dots, N\}$, of its components functioned in phase $i$. Both the system and its components are non-repairable during the mission, so $n \ge l_1 \ge l_2 \ge \dots \ge l_N \ge 0$ and the number of components that function at the beginning of phase $i$ is $m_i = l_{i-1},$ with $m_1=n$ --- so all components appear in all phases.  Subject to these constraints which do not apply in a non-PMS, the survival signature can then be applied without further modification for the mission completion time.

There are $\binom{m_i}{l_i}$ state vectors where precisely $l_i$ components function.  Because the random failure times of components in the same phase are independent and exchangeable, the survival signature is equal to:
\begin{equation}
  \Phi(l_1, \dots, l_N) = \left[ \prod_{i=1}^N \binom{m_i}{l_i}^{-1} \right] \sum_{\mathbf{X} \in \mathcal{S}} \varphi_S(\mathbf{X})
  \label{eq:survsig0}
\end{equation}
where $\mathcal{S}$ denotes the set of all possible state vectors for the whole system where $l_i$ components in phase $i$ are functioning.  This step is of the same form as the standard survival signature for a static system \citep{coolen2013generalizing}, but note one immediate subtle difference: as noted above, $m_i$ is not fixed across evaluations of $\Phi(\cdot)$, but rather is determined by $l_{i-1}$, since the maximum number of functioning components in the $i$th phase is determined by how many components completed phase $i-1$ still functioning.

A further subtlety arises as soon as we consider any time leading up to the mission completion time, because the structure of the system changes.  Although the standard survival signature can be used in computing the reliability of a static system at any point in its life \citep{coolen2013generalizing}, this is no longer true in this extension to PMSs.  Consequently, \eqref{eq:survsig0} is the survival signature which represents the probability that the whole mission completes successfully given that $l_i$ components are working in phase $i$.  For the survival function of a PMS, we must extend the survival signature to create a family of survival signatures which account for the temporally changing structure.  Let $\Phi_p(l_1, \dots, l_p)$ denote the survival signature of a PMS up to and including phase $p \le N$, which is the probability that the mission has not yet failed by phase $p$ given that $l_i$ components are working in phase $i \in \{1, \dots, p\}$.  Then,
\begin{equation}
  \Phi_p(l_1, \dots, l_p) = \left[ \prod_{i=1}^p \binom{m_i}{l_i}^{-1} \right] \sum_{\mathbf{X} \in \mathcal{S}} \prod_{i=1}^p \varphi_i(\mathbf{X})
  \label{eq:survsig0t}
\end{equation}

We define a function mapping mission time $t$ to the current phase
\begin{equation}
  \rho(t) : [0,\tau_{N+1}] \to \{1, \dots, N\}, \mbox{ as }  \rho(t) \triangleq \max\{ i \,:\, \tau_i < t \} \label{eq:currentphase}
\end{equation}

From \cref{eq:allphases} and \eqref{eq:survsig0t}, the reliability of the PMS at time $t$ can then be expressed pointwise as:
\begin{equation}
  R(t) = \sum_{l_1=0}^{m_1} \cdots \sum_{l_{\rho(t)}=0}^{m_{\rho(t)}} \left[ \Phi_{\rho(t)}(l_1, \dots, l_{\rho(t)}) \mathbb{P}\left( \bigcap_{i=1}^{\rho(t)} \left\{ C_i(t) = l_i \right\} \right) \right]
  \label{eq:survsig1}
\end{equation}
where $C_i(t)$ is the random variable denoting the number of components in phase $i$ which function at time $t \in [\tau_i, \tau_{i+1})$.  If $R(t)$ is being evaluated at $t \ge \tau_{i+1}$ then $C_i(t) \triangleq C_i(\tau_{i+1})$.  By the definition of $\rho(t)$, $R(t)$ will never be evaluated for $t < \tau_{i}$.

Because components are of the same type they share a common lifetime distribution as long as they all appear in all phases (and hence age together).  As a result, the sequential nature of a PMS means that components in the same phase have common conditional CDF, $F_i(t)$, for phase $i$, where conditioning is on the component having worked at the beginning of phase $i$.  That is, if the components have common CDF $F(t)$ and all components appear in every phase (in possibly different configurations), then the conditional CDF in phase $i$ is:
\begin{align}
  F_i(t) &= \mathbb{P}(T < t \,|\, \tau_i, \tau_{i+1}, T > \tau_i) \nonumber \\
  &= \frac{1}{1-F(\tau_i)} \int_{\tau_i}^{\min \{t, \tau_{i+1}\}} dF(z) \nonumber \\
  &= \frac{F(\min \{t, \tau_{i+1}\}) - F(\tau_i)}{1-F(\tau_i)} \label{eq:condcdf}
\end{align}
where $\tau_i$ is the start time of phase $i$ ($\tau_1 \triangleq 0$) and $T$ is the random variable representing component lifetime.

Proceeding with this conditional CDF, the last term in \cref{eq:survsig1} can be simplified as
\begin{align*}
  \mathbb{P}\left( \bigcap_{i=1}^{\rho(t)} \left\{ C_i(t) = l_i \right\} \right) &= \prod_{i=1}^{\rho(t)} \mathbb{P}\left( C_i(t) = l_i \right) \\
  &= \prod_{i=1}^{\rho(t)} \left[ \binom{m_i}{l_i} (R_i(t))^{l_i} (1-R_i(t))^{m_i-l_i} \right]
\end{align*}
where
\begin{equation}
  R_i(t) = 1-F_i(t) = \frac{1-F(\min \{t, \tau_{i+1}\})}{1-F(\tau_i)} \label{eq:comprel}
\end{equation}
is the reliability of the components at time $t$ in phase $i$.

Thus, \cref{eq:survsig1} can be rewritten pointwise in $t$ as
\begin{align}
  R(t) &= \sum_{l_1=0}^{m_1} \cdots \sum_{l_{\rho(t)}=0}^{m_{\rho(t)}} \left\{ \Phi_{\rho(t)}(l_1, \dots, l_{\rho(t)}) \phantom{\prod_{i=1}^{\rho(t)} \left[ \binom{m_i}{l_i} (R_i(t))^{l_i} (1-R_i(t))^{m_i-l_i} \right]} \right. \nonumber \\
  &\qquad\qquad\qquad\qquad \times \left. \prod_{i=1}^{\rho(t)} \left[ \binom{m_i}{l_i} (R_i(t))^{l_i} (1-R_i(t))^{m_i-l_i} \right] \right\} \label{eq:survsigPMS1a}
\end{align}

Since in the general case (see special case exception in the sequel) every component appears in every phase, this can be written
\begin{align}
  R(t) &= \sum_{l_1=0}^{l_0} \cdots \sum_{l_{\rho(t)}=0}^{l_{\rho(t)-1}} \left\{ \Phi_{\rho(t)}(l_1, \dots, l_{\rho(t)}) \phantom{\prod_{i=1}^{\rho(t)} \left[ \binom{m_i}{l_i} (R_i(t))^{l_i} (1-R_i(t))^{m_i-l_i} \right]} \right. \nonumber \\
  &\qquad\qquad\qquad\qquad \times \left. \prod_{i=1}^{\rho(t)} \left[ \binom{l_{i-1}}{l_i} (R_i(t))^{l_i} (1-R_i(t))^{l_{i-1}-l_i} \right] \right\} \label{eq:survsigPMS1b}
\end{align}
where we define $l_0 \triangleq n$.  Writing in this final form stresses the sequential dependence in the computation, in stark contrast to the standard survival signature for a static system.

\subsubsection{Special case: Exponentially distributed component lifetime}

There are two simplifications that arise when components are Exponentially distributed.  Firstly, $F_i(t) \equiv F(t) \ \forall\, i$, so that $R_i(t) = R(t) = 1-F(t-\tau_i) \ \forall i$.

The second simplification is that not all components need to appear in all phases.  It may be that some components appear only in later phases (but continue to appear after the first phase they are in).  In this case, one should be careful not to use \eqref{eq:survsigPMS1b}, but instead \eqref{eq:survsigPMS1a} where now $m_i=l_{i-1}+m_i^\star$ where $m_i^\star$ is the number of components appearing in the system for the first time at phase $i$.

\subsubsection{Modelling constraints}

\begin{figure}
  \centering
  \includegraphics[width=0.8\textwidth]{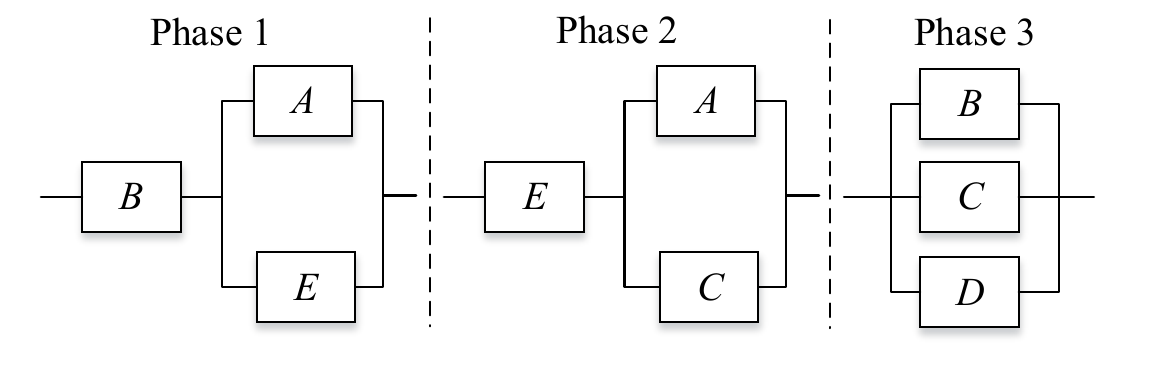}
  \caption{A PMS with multiple types of components.}
  \label{fig:pms2}
\end{figure}

Note that considerable care is required in the specification of --- and implicit assumptions made for --- $F_i(t)$.  In particular, when a component is not present in a phase, then whether ageing continues (i.e.\ time passes) or not is crucial in determining whether the assumption of identical component lifetime distribution still holds in all phases.  For example, in \cref{fig:pms1} each component appears in all phases and therefore experiences the same wear, but in \cref{fig:pms2} each component is in precisely 2 of the 3 phases.  Consequently, even though one might assume all components are of the same type initially, if component $C$ is considered not to `age' during phase 1 (where it is not present) then it will in fact not have identical conditional lifetime distribution to $A$ and $E$ during phase 2, since the latter will have already experienced wear from phase 1.

This imposes rather unattractive modelling strictures: all components of similar type must appear in the same phases; or all components must have constant failure rate (Exponentially distributed lifetime).  These modelling strictures severely limit applicability to real world systems, thus motivating the novel methodological extension of survival signatures hereinafter.

\subsection{PMSs with different components in different phases}
\label{sec:pms.diff}

Most practical PMSs for which the reliability is modelled consist of heterogeneous component types both within and between phases. Therefore, a more interesting challenge is to extend the methodology of survival signatures to this more general setting.

We now consider this setting in generality and show that the problem again simplifies in the special case of Exponentially distributed lifetimes, which is the only case that most of the literature has addressed to date.  The only constraint we impose is that components of the same type appear in the same phases (since then the conditional CDFs within phases remain in agreement).  However, note that this does not limit the scenarios that can be modelled, since components of the same physical type can still be split into multiple `meta-types'.

\begin{definition}{(Meta-type)}
  Components are defined to be of the same \emph{meta-type} when they are of the same physical type and appear in the same phases.
\end{definition}

Let there be a total of $K$ different meta-types of component.  We take the multi-type, multi-phase survival signature to be denoted by the function $\Phi(l_{11}, \dots, l_{1K}, \dots, l_{N1}, \dots, l_{NK})$, the probability that the system functions given that precisely $l_{ik}$, components of type $k$ function in phase $i$.  That is,
\[ \Phi(l_{11}, \dots, l_{1K}, \dots, l_{N1}, \dots, l_{NK}) = \left[ \prod_{i=1}^N \prod_{k=1}^{K} \binom{m_{ik}}{l_{ik}}^{-1} \right] \sum_{\mathbf{X} \in \mathcal{S}} \varphi_S(\mathbf{X}) \]
where $\mathcal{S}$ denotes the set of all possible state vectors for the whole system.  Not all component types need necessarily appear in all phases, so we admit the possibility that $m_{ik}=0$ when a component type is absent from a phase and observe the standard definition that $\binom{0}{0} \triangleq 1$ --- this simplifies notation versus having varying numbers of $l_{i\cdot}$ for each phase.

As before, the above survival signature is only applicable to the full mission time and we define a family of survival signatures corresponding the successive phases of the mission.  Let $\Phi_p(l_{11}, \dots, l_{1K}, \dots, l_{p1}, \dots, l_{pK})$ denote the survival signature of a PMS up to and including phase $p \le N$, which is the probability that the mission has not yet failed by phase $p$ given that $l_{ik}$ components of type $k$ are working in phase $i \in \{1, \dots, p\}$.  Then,
\begin{equation}
  \Phi_p(l_{11}, \dots, l_{1K}, \dots, l_{p1}, \dots, l_{pK}) = \left[ \prod_{i=1}^p \prod_{k=1}^{K} \binom{m_{ik}}{l_{ik}}^{-1} \right] \sum_{\mathbf{X} \in \mathcal{S}} \prod_{i=1}^p \varphi_i(\mathbf{X})
  \label{eq:survsig0t2}
\end{equation}

We retain the definition of $\rho(t)$ given in \eqref{eq:currentphase}.  It then follows from \cref{eq:allphases} and \cref{eq:survsig0t2} that the reliability of the PMS can be characterised as:
\begin{align}
  R(t) &= \sum_{l_{11}=0}^{m_{11}} \cdots \sum_{l_{\rho(t),K}=0}^{m_{\rho(t),K}} \left[ \Phi_{\rho(t)}(l_{11}, \dots, l_{1K}, \dots, l_{\rho(t),1}, \dots, l_{\rho(t),K}) \vphantom{\mathbb{P}\left( \bigcap_{i=1}^N \bigcap_{k=1}^{K_i} \left\{ C_{ik}(t) = l_{ik} \right\} \right)} \right. \nonumber \\
  & \qquad\qquad\qquad\qquad\qquad \left. \times \mathbb{P}\left( \bigcap_{i=1}^{\rho(t)} \bigcap_{k=1}^{K} \left\{ C_{ik}(t) = l_{ik} \right\} \right) \right] \label{eq:survsig2}
\end{align}
where $C_{ik}(t)$ is the random variable denoting the number of components of type $k$ in phase $i$ which function at time $t \in [\tau_i, \tau_{i+1})$.  In the same vein as \cref{sec:pms.same}, if $R(t)$ is being evaluated at $t \ge \tau_{i+1}$ then $C_{ik}(t) \triangleq C_{ik}(\tau_{i+1})$.  By the definition of $\rho(t)$, $R(t)$ will never be evaluated for $t < \tau_{i}$.

We can simplify, by defining that $\mathbb{P}\left( C_{ik}(t) = 0 \right) = 1$ when $m_{ik}=0$.
\begin{align*}
  \mathbb{P}\left( \bigcap_{i=1}^{\rho(t)} \bigcap_{k=1}^{K} \left\{ C_{ik}(t) = l_{ik} \right\} \right) &= \prod_{i=1}^{\rho(t)} \prod_{k=1}^{K} \mathbb{P}\left( C_{ik}(t) = l_{ik} \right) \\
  &= \prod_{i=1}^{\rho(t)} \prod_{k=1}^{K} \left[ \binom{m_{ik}}{l_{ik}} (R_{ik}(t))^{l_{ik}} (1-R_{ik}(t))^{m_{ik}-l_{ik}} \right]
\end{align*}
with
\begin{equation}
  R_{ik}(t) = \frac{1-F_k(\min \{t, \tau_{i+1}\})}{1-F_k(\tau_i)} \label{eq:comprel2}
\end{equation}
where $F_k(\cdot)$ is the CDF of the component lifetime distribution for the meta-type $k$.

Consequently, for any time $t$ during the mission, we have the reliability of the system characterised by:

\begin{align}
  R(t) &= \sum_{l_{11}=0}^{m_{11}} \cdots \sum_{l_{\rho(t),K}=0}^{m_{\rho(t),K}} \left\{ \Phi_{\rho(t)}(l_{11}, \dots, l_{1K}, \dots, l_{\rho(t),1}, \dots, l_{\rho(t),K}) \vphantom{\prod_{i=1}^{\rho(t)} \prod_{k=1}^{K} \left[ \binom{m_{ik}}{l_{ik}} (R_{ik}(t))^{l_{ik}} (1-R_{ik}(t))^{m_{ik}-l_{ik}} \right]} \right. \nonumber \\
  & \qquad\qquad\ \left. \times \prod_{i=1}^{\rho(t)} \prod_{k=1}^{K} \left[ \binom{m_{ik}}{l_{ik}} (R_{ik}(t))^{l_{ik}} (1-R_{ik}(t))^{m_{ik}-l_{ik}} \right] \right\} \label{eq:survsigPMS2}
\end{align}
where $m_{ik} = l_{jk}$ for $j = \max \{ j : j < i, m_{jk} > 0 \}$.  That is, $m_{ik}$ is the number components which were working in the most recent preceding phase where this component meta-type appears.

\subsubsection{Special case: Exponential component lifetimes}

Exponentially distributed component lifetimes again provide simplifications.  Now, the $R_{ik}(t) \equiv R_k(t)$ due to the memoryless property of the Exponential distribution.

Furthermore, we can relax the definition of a meta-type of component.  The definition of component meta-types serves two purposes: (i) to ensure that $m_{ik}$ can be determined without tracking the individual functioning status of all components; and (ii) to ensure that the conditional CDFs of all components of the same meta-type in a phase are the same.  The second purpose is made entirely redundant by the memoryless nature of the Exponential distribution.  The first purpose remains, but can be achieved with a weaker definition of meta-type.

\begin{definition}{(Exponential meta-type)}
  Components are defined to be of the same \emph{exponential meta-type} when they are of the same Exponentially distributed physical type, and if once any pair of components of the same \emph{exponential meta-type} appear in a phase together, they both appear in all subsequent phases where either component appears.
\end{definition}

In other words, components of the same exponential meta-type may first appear in the system at different phases, but thereafter should appear whenever at least one such exponential meta-type component appears.  This definition enables the determination of $m_{ik}$ as $m_{ik} = l_{jk} + m_{ik}^\star$ for $j = \max \{ j : j < i, m_{jk} > 0 \}$, where $m_{ik}^\star$ is the number of components of exponential meta-type $k$ appearing for the first time in phase $i$.

The benefits of Exponential component lifetimes can be mixed in a system containing both meta-type and exponential meta-types since a crucial feature of survival signatures is the factorisation of such types so that they do not interact.

\section{Numerical examples}
\label{sec:examples}

\subsection{Example 1}

We first consider the PMS shown in \cref{fig:pms1}. The duration of each phase is taken to be 10 hours, and the failure rate of each component in each phase is $10^{-4}$/hour.

The survival signatures of this PMS can be obtained using \cref{eq:survsig0}.  The elements of the survival signature which are non-zero are shown in \cref{tab:pms1.survsig} --- that is, rows where $\Phi(l_1)=0, \Phi(l_1, l_2)=0$ and $\Phi(l_1, l_2, l_3)=0$ are omitted. The table is grouped into a nested sequence of phases, with just the first phase shown, followed by the first two phases together and finally all phases --- this helps emphasise and clarify the sequential dependence of phases, where $m_k$ depends on $l_{k-1}$.

\begin{table}
  \centering\renewcommand{\arraystretch}{1.25}
  \begin{tabular}{ccccccccc}
    \hline
    \multicolumn{2}{c}{First phase} & \multicolumn{3}{c}{Phase 1+2} & \multicolumn{4}{c}{All Phases} \tabularnewline
    \multicolumn{2}{c}{$0 \le t \le 10$} & \multicolumn{3}{c}{$10 < t \le 20$} & \multicolumn{4}{c}{$20 < t \le 30$} \tabularnewline
    \hline 
    $l_1$ & $\Phi(l_1)$ & $l_1$ & $l_2$ & $\Phi(l_1, l_2)$ & $l_1$ & $l_2$ & $l_3$ & $\Phi(l_{1},l_{2},l_{3})$ \tabularnewline
    \hline 
    \hline
    3 & 1 & 3 & 1 & 1 & 3 & 2 & 2 & $\frac{2}{3}$ \tabularnewline
     & & 3 & 2 & 1 & 3 & 3 & 2 & $\frac{2}{3}$ \tabularnewline
     & & 3 & 3 & 1 & 3 & 3 & 3 & 1 \tabularnewline
    \hline 
  \end{tabular}
  \caption{Survival signature of the PMS shown in \cref{fig:pms1}}
  \label{tab:pms1.survsig}
\end{table}

\begin{figure}
  \centering
  \includegraphics[width=0.8\textwidth]{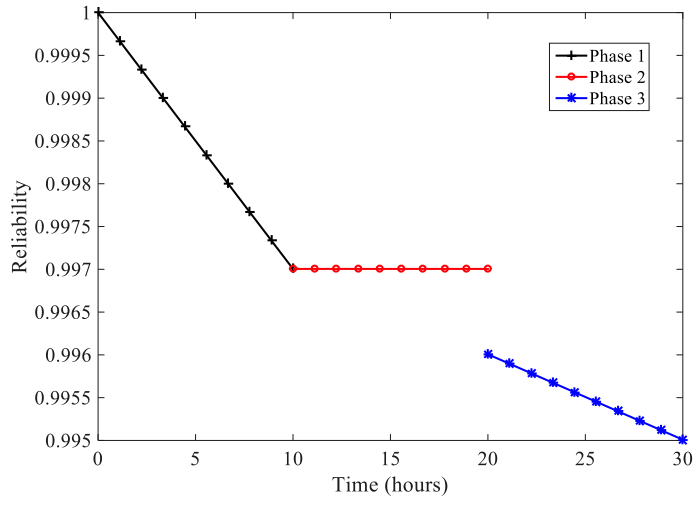}
  \caption{Reliability of the PMS in example 1.}
  \label{fig:pms1.R}
\end{figure}

We can obtain the conditional reliability of components using the conditional failure rate of the component in each phase.  \Cref{eq:survsigPMS1a} then renders the reliability of the PMS as a whole. The results are shown in \cref{tab:pms1.R} and \cref{fig:pms1.R}.  These results concord with those found using an independent method in \citep{zang1999bdd}.

Of note is the jump discontinuity in the reliability function at $t=20$, as shown in \cref{fig:pms1.R}. This occurs because a failure of component $A$ during phase 2 does not necessarily cause failure of the system at that point, so long as at least one of components $B$ or $C$ work.  However, in this situation the PMS will fail instantaneously upon commencing phase 3 at $t=20^+$.  Consequently, the size of the jump discontinuity in fact corresponds to the probability of the event $\{ A$ fails in phase 2, but the system still functions$\}$.

\begin{table}
  \centering\renewcommand{\arraystretch}{1.25}
  \begin{tabular*}{0.85\textwidth}{@{\extracolsep{\fill}}ccccccc@{\extracolsep{\fill}}}
    \hline
    $t$ & $0$ & $10^{-}$ & $10^{+}$ & $20^{-}$ & $20^{+}$ & $30$ \tabularnewline
    $R$ & 1 & 0.99700 & 0.997700 & 0.997700 & 0.99601 & 0.99501 \tabularnewline
    \hline 
  \end{tabular*}
  \caption{Reliability of the PMS in example 1}
  \label{tab:pms1.R}
\end{table}

\subsection{Example 2}

For the PMS shown in \cref{fig:pms2}, phases 1, 2 and 3 last for 10, 90 and 100 hours respectively.  All components in each phase are of the same type and the lifetime distribution of these components follows a two-parameter Weibull distribution.  \Cref{tab:pms2.pars} summarises the distribution information of the components in each phase.

\begin{table}
  \centering\renewcommand{\arraystretch}{1.25}
  \begin{tabular*}{0.75\textwidth}{@{\extracolsep{\fill}}lccc@{\extracolsep{\fill}}}
    \hline
    Parameter & Phase 1 & Phase 2 & Phase 3 \tabularnewline\hline 
    Scale & 250 & 1000 & 300 \tabularnewline
    Shape & 2.6 & 3.2 & 2.6 \tabularnewline
    \hline 
  \end{tabular*}
  \caption{Conditional distribution information of the components in each phase}
  \label{tab:pms2.pars}
\end{table}

As described in \cref{sec:pms.diff}, if some components of the same type appear in a phase and also appear in some subsequent phases --- but not simultaneously --- then these components should be considered as different types of component. For this example, this means that despite the fact they all share a common failure rate within phases, components $A$ and $E$ need to be labelled as type 1 and the remainder as type 2, because ageing will have been different.

\begin{figure}
  \centering
  \includegraphics[width=0.8\textwidth]{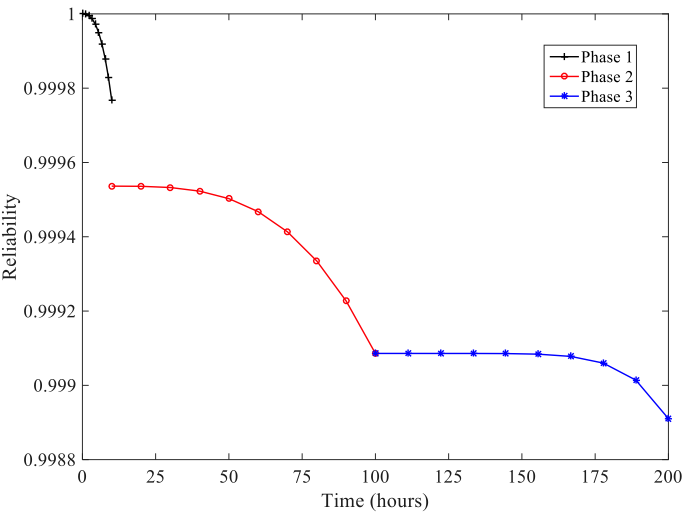}
  \caption{Reliability of the PMS in example 2.}
  \label{fig:pms2.R}
\end{figure}

The survival signatures of this PMS are shown in \cref{tab:pms2.survsig}, with rows where $\Phi(l_{11},l_{12})=0, \Phi(l_{11},l_{12},l_{21},l_{22})=0$ and $\Phi(l_{11},l_{12},l_{21},l_{22},l_{32})=0$ suppressed.  The reliability of the PMS is shown in \cref{tab:pms2.R} and \cref{fig:pms2.R}.

We again see a jump discontinuity in the reliability curve depicted in \cref{fig:pms2.R}, at $t=10$.  In this instance, if component $E$ fails during phase 1 the system will still function, but instantaneous failure will occur once phase 2 commences. This is evident in \cref{tab:pms2.R}, which shows the jump discontinuity is of size $2.3 \times 10^{-4}$.  Indeed, this should correspond to the probability that the system survives phase 1 but with component $E$ failing during that phase.  That is:
\begin{align*}
   & \mathbb{P}(A, B \mbox{ function} \cap E \mbox{ fails in phase 1}) \\
   & \quad= \mathbb{P}(E \mbox{ fails in phase 1}) \mathbb{P}(A, B \mbox{ function} \,|\, E \mbox{ fails in phase 1}) \\
   & \quad = \int_0^{10} b a^{-b} t^{b-1} e^{-(t/a)^b}\,dt \left(1 - \int_0^{10} b a^{-b} t^{b-1} e^{-(t/a)^b}\,dt \right)^2 \\
  & \quad \approx 2.3 \times 10^{-4}  \ \ \mbox{for } a=250, b=2.6
\end{align*}
as required.  Hence, PMS can exhibit jump discontinuities where probability mass from non-critical failures in one phase accumulate onto phase change boundaries when the system layout switches.

\begin{table}
  \centering\renewcommand{\arraystretch}{1.25}
  \begin{tabular}{llllllllllllll}
    \hline
    \multicolumn{3}{l}{The first phase} & \multicolumn{5}{l}{The first two phases} & \multicolumn{6}{l}{All phases}         \tabularnewline
    \hline 
    $l_{11}$        & $l_{12}$       & $\Phi_1$       & $l_{11}$   & $l_{12}$   & $l_{21}$   & $l_{22}$  & $\Phi_{12}$  & $l_{11}$ & $l_{12}$ & $l_{21}$ & $l_{22}$ & $l_{32}$ & $\Phi_S$  \tabularnewline
    \hline
    \hline
    1           & 1          & 1        & 1      & 1      & 1      & 1     & 1/2   & 1    & 1    & 1    & 1    & 1    & 1/2 \tabularnewline
    2           & 1          & 1        & 2      & 1      & 1      & 1     & 1/2   & 1    & 1    & 1    & 1    & 2    & 1/2 \tabularnewline
                &            &          & 2      & 1      & 2      & 0     & 1     & 1    & 1    & 1    & 1    & 3    & 1/2 \tabularnewline
                &            &          & 2      & 1      & 2      & 1     & 1     & 2    & 1    & 1    & 1    & 1    & 1/2 \tabularnewline
                &            &          &        &        &        &       &       & 2    & 1    & 1    & 1    & 2    & 1/2 \tabularnewline
                &            &          &        &        &        &       &       & 2    & 1    & 1    & 1    & 3    & 1/2 \tabularnewline
                &            &          &        &        &        &       &       & 2    & 1    & 2    & 0    & 1    & 1   \tabularnewline
                &            &          &        &        &        &       &       & 2    & 1    & 2    & 0    & 2    & 1   \tabularnewline
                &            &          &        &        &        &       &       & 2    & 1    & 2    & 1    & 1    & 1   \tabularnewline
                &            &          &        &        &        &       &       & 2    & 1    & 2    & 1    & 2    & 1   \tabularnewline
                &            &          &        &        &        &       &       & 2    & 1    & 2    & 1    & 3    & 1  \tabularnewline
    \hline
  \end{tabular}
  \caption{Survival signature of the PMS shown in \cref{fig:pms2}}
  \label{tab:pms2.survsig}
\end{table}

\begin{table}
  \centering\renewcommand{\arraystretch}{1.25}
  \begin{tabular*}{0.85\textwidth}{@{\extracolsep{\fill}}ccccccc@{\extracolsep{\fill}}}
    \hline
    $t$ & $0$ & $10^{-}$ & $10^{+}$ & $100^{-}$ & $100^{+}$ & $200$ \tabularnewline
    $R$ & 1 & 0.999768 & 0.999536 & 0.999086 & 0.999086 & 0.998910 \tabularnewline
    \hline 
  \end{tabular*}
  \caption{Reliability of the PMS in example 2.}
  \label{tab:pms2.R}
\end{table}

\subsection{Example 3}

\begin{figure}
  \centering
  \includegraphics[width=1\textwidth]{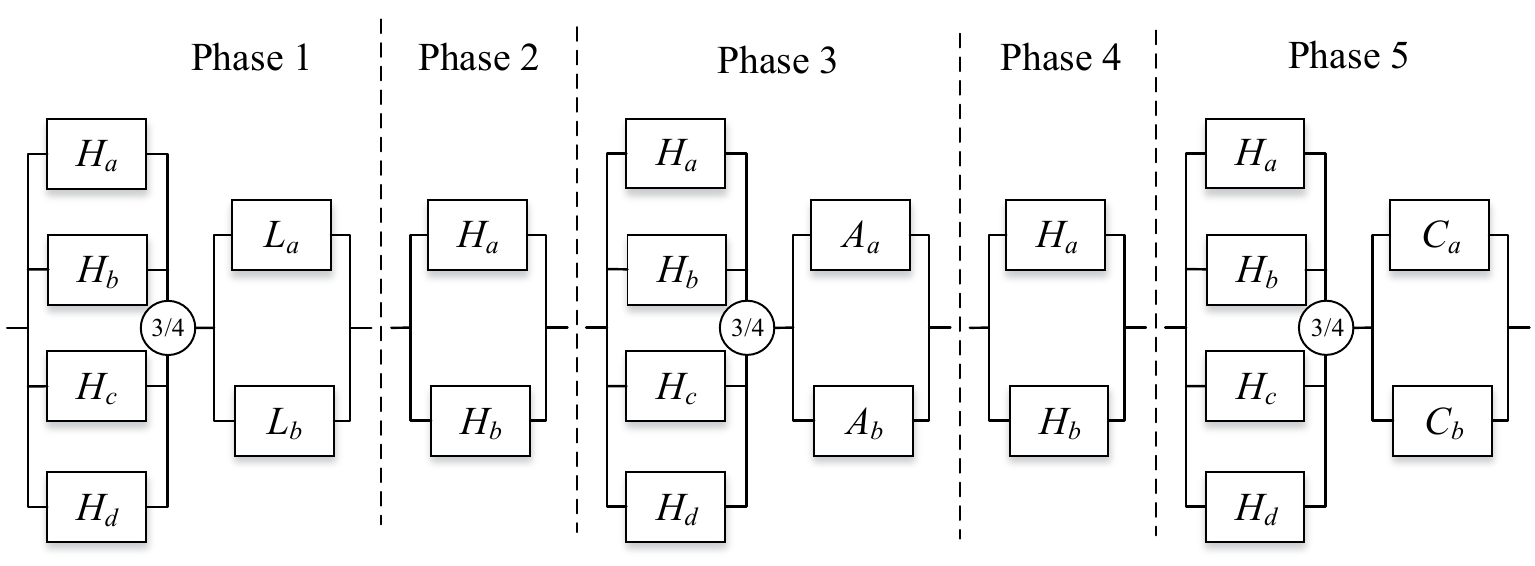}
  \caption{Reliability block diagram of the space application.}
  \label{fig:pms3}
\end{figure}

In this final example, we replicate the space application mission discussed by Zang \citep{zang1999bdd} and Mural \citep{mural1999dependability}.  This example includes the full complexity of real-world PMSs, where there is now heterogeneity of component types within phases.  This means that multiple component types arise necessarily and not merely as a side effect of identical components appearing in differing phases.  There are five phases involved in this space mission: launch is the first phase, followed by Hibern.1, Asteroid, Hibern.2, and finally Comet.  The reliability block diagram is shown in \cref{fig:pms3}. The five phases last for 48, 17520, 672, 26952 and 672 hours, respectively. The failure rates of the components in each phase are given in \cref{tab:pms3.lambda}.

\begin{table}
  \centering\renewcommand{\arraystretch}{1.25}
  \begin{tabular}{llllll}
    \hline
                   & Phase1  & Phase 2 & Phase 3 & Phase 4 & Phase 5 \tabularnewline
    \hline
    $H_a$, $H_b$, $H_c$, $H_d$ & $10^{-5}$  & $10^{-6}$  & $10^{-5}$  & $10^{-6}$  & $10^{-5}$ \tabularnewline
    $L_a$, $L_b$         & $5 \times 10^{-5}$  & 0       & 0       & 0       & 0      \tabularnewline
    $A_a$, $A_b$         & 0       & 0       & $10^{-5}$  & 0       & 0      \tabularnewline
    $C_a$, $C_b$         & 0       & 0       & 0       & 0       & $10^{-4}$ \tabularnewline
    \hline
  \end{tabular}
  \caption{Failure rates of the components.}
  \label{tab:pms3.lambda}
\end{table}

As shown in \cref{tab:pms3.types}, in order to calculate the reliability of the PMS, the 4 `real' component types must be divided into 5 types when using the methodology presented in this paper.  That is, although $H_a, H_b, H_c,$ and $H_d$ have homogeneous failure rates throughout all phases, because they do not always appear together they will exhibit different ageing.  Consequently, these are split into two `pseudo' types.

\begin{figure}
  \centering
  \includegraphics[width=0.8\textwidth]{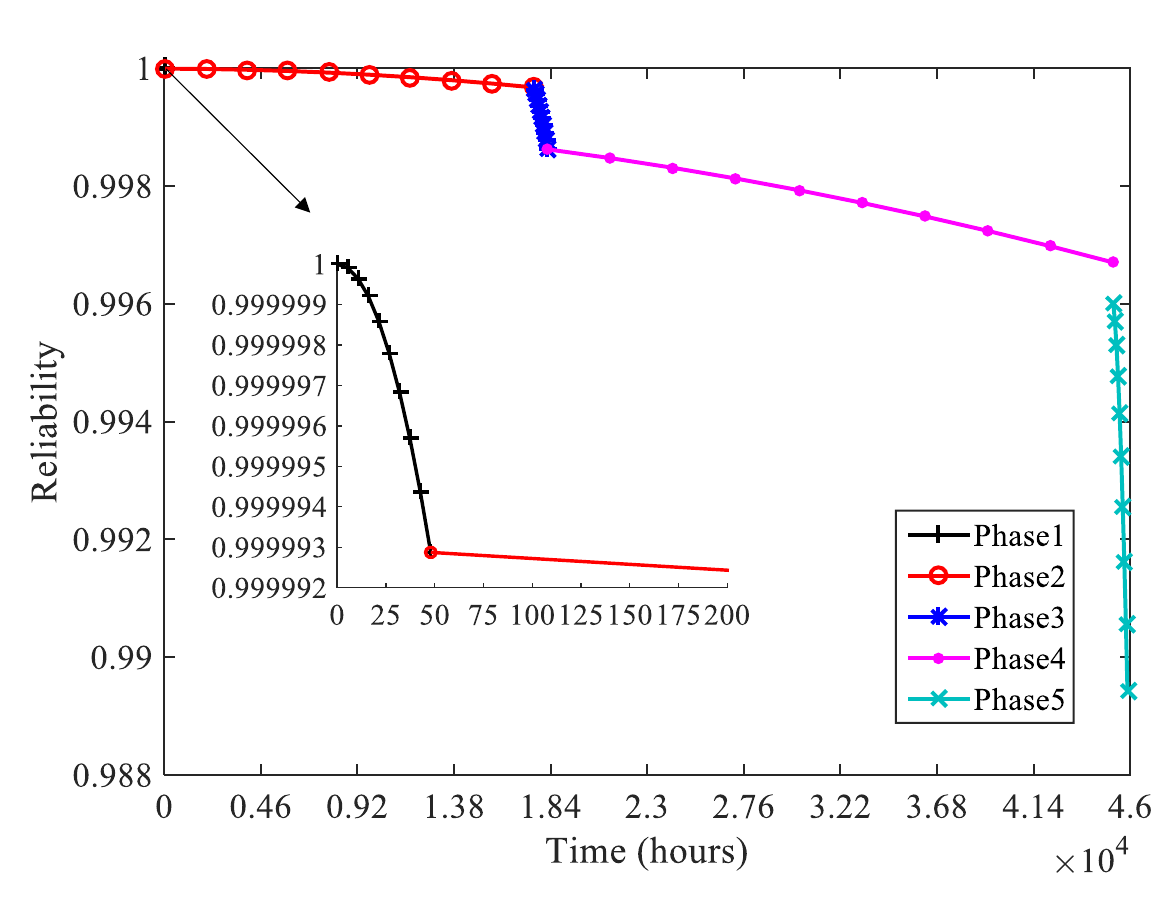}
  \caption{Reliability of the PMS in example 3, with inset graph providing blown-up detail of first 200 hours of operation.}
  \label{fig:pms3.R}
\end{figure}

The result of analysing the reliability of this PMS is shown in \cref{tab:pms3.R} and \cref{fig:pms3.R}. The results found using the new methodology we have presented in this paper are in agreement with the entirely independent method in \citep{zang1999bdd}. 

\begin{table}
  \centering\renewcommand{\arraystretch}{1.25}
  \begin{tabular*}{0.85\textwidth}{@{\extracolsep{\fill}}ccccc@{\extracolsep{\fill}}}
    \hline
    Type 1 & Type 2 & Type 3 & Type 4 & Type 5 \tabularnewline
    $H_a$, $H_b$ & $H_c$, $H_d$ & $L_a$, $L_b$ & $A_a$, $A_b$ & $C_a$, $C_b$ \tabularnewline
    \hline 
  \end{tabular*}
  \caption{Types of components in example 3.}
  \label{tab:pms3.types}
\end{table}

\begin{table}
  \centering\renewcommand{\arraystretch}{1.25}
  \resizebox{0.9\textwidth}{!}{%
  \begin{tabular}{ccccccccccc}
    \hline
    $t$ & $0$ & $48^{-}$ & $48^{+}$ & $17568^{-}$ & $17568^{+}$ & $18240^{-}$ & $18240^{+}$ & $45192^{-}$ & $45192^{+}$ & $45864$ \tabularnewline
    $R$ & 1 & 0.99999 & 0.99999 & 0.99968 & 0.99964 & 0.99862 & 0.99862 & 0.99670 & 0.99600 & 0.98943 \tabularnewline
    \hline 
  \end{tabular}}
  \caption{Reliability of the PMS in example 3.}
  \label{tab:pms3.R}
\end{table}

\section{Conclusion}
\label{sec:conclusion}

Computing the reliability of a PMS is considerably more complex than that of a non-PMS, due to the variation in system structure between phases and the dependencies between component failures in different phases.  Consequently, reliability analysis of PMSs has become one of the most challenging topics in the field of system reliability evaluation and maintenance engineering in recent decades.  Despite some progress towards efficient and effective methods for measuring the reliability of PMS, it is still difficult to analyze large systems without considerable computational expense and even where it is possible, many methods fail to convey intuition about the reliability of the system.

In this paper, a new and efficient method for reliability analysis of PMS is proposed using survival signature. Signatures have been proven to be an efficient method for estimating the reliability of systems. A new kind of survival signature is derived to represent the structure function of the PMS. Then the proposed survival signature is applied to calculate the reliability of the PMS. Reliability analysis of a system using signatures could separate the system structure from the component probabilistic failure distribution. Therefore, the proposed approach is easy to be implemented in practice and has high computational efficiency.

Note that reliability analysis of PMSs with multiple failure mode components is not studied in this paper. In practice the components may perhaps have more than one failure mode. In ongoing work, the authors are considering component importance analysis, extending work such as \cite{feng2016imprecise, eryilmaz2018marginal} to PMSs.
\section*{Acknowledgements}

The authors gratefully acknowledge the support of National Natural Science Foundation of China (51575094), China Postdoctoral Science Foundation (2017M611244), China Scholarship Council (201706085013) and Fundamental Research Funds for the Central Universities (N160304004).

This work was performed whilst the first author was a visitor at Durham University.

\section*{References}

\bibliographystyle{elsarticle-num}
\bibliography{references}

\end{document}